\documentclass[lettersize,journal]{IEEEtran}
\PassOptionsToPackage{hyphens}{url}

\usepackage{makecell}
\usepackage{longtable}
\usepackage{booktabs}
\usepackage{float}
\usepackage{amsmath,amsfonts}
\usepackage{algorithmic}
\usepackage{algorithm}
\usepackage{array}
\usepackage[caption=false,font=normalsize,labelfont=sf,textfont=sf]{subfig}
\usepackage{textcomp}
\usepackage{stfloats}
\usepackage{url}
\usepackage{verbatim}
\usepackage{graphicx}
\usepackage{cite}
\usepackage{multirow} 
\usepackage[colorlinks,linkcolor=red,anchorcolor=green,citecolor=blue]{hyperref}
\usepackage{cleveref}

\usepackage[table]{xcolor}
\crefname{figure}{fig}{figures}
\Crefname{figure}{Fig}{Figures}
\begin{document}

\title{Real World Assets on-Chain Assistance Low-Altitude Computility Networks: Architecture, Methodology, and Challenges}

\author{Haoxiang Luo, Ruichen Zhang, Yinqiu Liu, \\Gang Sun,~\IEEEmembership{Senior Member,~IEEE}, Hongfang Yu,~\IEEEmembership{Senior Member,~IEEE}, \\ and Zhu Han,~\IEEEmembership{Fellow,~IEEE}
\thanks{H. Luo, G. Sun, and H. Yu are with the School of Information and Communication Engineering, University of Electronic Science and Technology of China, Chengdu 611731, China (e-mail: lhx991115@163.com; \{gangsun, yuhf\}@uestc.edu.cn).}
\thanks{R. Zhang and Y. Liu are with the College of Computing and Data Science, Nanyang Technological University, Singapore 639798 (e-mail: ruichen.zhang@ntu.edu.sg; yinqiu001@e.ntu.edu.sg). }
\thanks{Z. Han is with the Department of Electrical and Computer Engineering, University of Houston, Houston, TX 77004, USA (e-mail: zhan2@uh.edu).}
\thanks{The corresponding author: Gang Sun.}

}



\maketitle

\begin{abstract}

Low-altitude airspace is becoming a new frontier for smart city services and commerce. Networks of drones, electric Vertical Takeoff and Landing (eVTOL) vehicles, and other aircraft, termed Low-Altitude Economic Networks (LAENets), promise to transform urban logistics, aerial sensing, and communication. A key challenge is how to efficiently share and trust the computing utility, termed “computility”, of these aerial devices. We propose treating the computing power on aircraft as tokenized Real-World Assets (RWAs) that can be traded and orchestrated via blockchain. By representing distributed edge computing resources as blockchain tokens, disparate devices can form Low-Altitude Computility Networks (LACNets), collaborative computing clusters in the sky. We first compare blockchain technologies, non-fungible tokens (NFTs), and RWA frameworks to clarify how physical hardware and its computational output can be tokenized as assets. Then, we present an architecture using blockchain to integrate aircraft fleets into a secure, interoperable computing network. Furthermore, a case study models an urban logistics LACNet of delivery drones and air-taxis. Simulation results indicate improvements in task latency, trust assurance, and resource efficiency when leveraging RWA-based coordination. Finally, we discuss future research directions, including AI-driven orchestration, edge AI offloading and collaborative computing, and cross-jurisdictional policy for tokenized assets.

\end{abstract}


\section{Introduction} \label{sec-I}

\IEEEPARstart {U}{nmanned} Aerial Vehicles (UAVs), ranging from delivery drones to passenger electric Vertical Takeoff and Landing (eVTOL) aircraft, are driving the rise of the Low-Altitude Economy (LAE) \cite{wang2025toward}. In urban areas, these platforms enable key applications such as on-demand parcel delivery, aerial surveillance, emergency response, and urban air mobility. Collectively, dense networks of UAVs and eVTOLs form Low-Altitude Economic Networks (LAENets), which operate in the lower airspace to provide logistics, communication, and sensing services \cite{qin2025latency}. Governments and industry stakeholders are heavily investing in this vision. For example, the Civil Aviation Administration of China has outlined plans to “vigorously develop a low-altitude new economy driven by new intelligent unmanned aerial vehicles”, expanding drone logistics routes and urban air mobility services\footnote{https://www.unmannedairspace.info/emerging-regulations/china-to-accelerate-low-altitude-economy-of-long-distance-cargo-drones-and-urban-evtols}. LAENets hold the promise of enhanced efficiency and connectivity in cities, but they also face complex technical and operational challenges.

Unlike traditional isolated drone fleets, large-scale LAENets involve dense autonomous aerial nodes coordinating in real time within constrained airspace \cite{luo2024escm}. Ensuring efficient air traffic management, avoiding collisions, and preventing interference with ground networks requires robust coordination. Moreover, security risks are heightened, and low-altitude aircraft are vulnerable to cyberattacks and signal spoofing from the ground \cite{luo2025convergence}. As a result, the central issue is trust: \emph{ how can multiple stakeholders, such as drone operators, service providers, and regulators, reliably cooperate and share resources in an open airspace economy?}

Another emerging challenge is that drones, eVTOLs, and other aircraft carry significant computing resources on board (CPUs, GPUs, sensors) that often remain underutilized or siloed. Each aerial vehicle is effectively a flying edge-computer that could process data or run AI tasks locally. If harnessed collectively, these distributed computers aloft could form a powerful edge cloud to support the network’s needs, for example, analyzing sensor data, optimizing routes, or hosting local services. However, orchestrating this “computility” (computing utility) among many independent devices requires new approaches to resource sharing and incentive alignment. Drone operators need assurance they will be compensated for contributing compute power, and task offloading must occur in a secure, verifiable manner, given the high stakes, such as safety and privacy in aerial operations \cite{cai2025secure}.

This article explores a solution at the intersection of blockchain and Internet of Things (IoT). It regards the computing capabilities of drones, eVTOL vehicles, and other flying devices as tokenized assets on the blockchain. By leveraging Real-World Asset (RWA) tokenization, we can represent the computility or services of each device as a digital token that is tradable, interoperable, and anchored in the real asset’s performance \cite{chen2024exploring}. In essence, the physical capabilities of drones become “compute coins” that can be exchanged or rented in an open marketplace, enabling LACNets on top of the physical LAENets. Such tokens allow fractional ownership or usage rights of aerial compute resources, lowering the barrier for participants to access and trust these resources\footnote{https://coinpedia.org/information/exabits-rwa-approach-to-tokenized-gpu-power-democratises-ai-cloud-computing-ownership}. Additionally, blockchain smart contracts manage the orchestration, automatically matching computing tasks to available drone nodes, handling micro-payments, logging usage, and enforcing the rules without centralized control.

As far as we know, this is the first work that uses RWA to construct LACNets. This approach of leveraging RWA to tokenize the computing resources of aircraft has the following specific contributions to the construction of LACNets:

\begin{itemize} 

\item We introduce the novel concept of LACNets, built upon existing LAENets. The core innovation is the proposal to treat the underutilized onboard computing power of aerial vehicles like drones and eVTOLs as a tradable utility, termed ``computility." This is achieved by tokenizing this computing capacity as an RWA on a blockchain. This approach transforms siloed, physical hardware into liquid, interoperable digital assets, creating a collaborative, sky-based computing cluster.

\item We detail a technical framework for implementing LACNets. It is a hybrid blockchain design that uses a permissioned chain for secure identity management among trusted entities and a permissionless chain for managing the tokenized RWAs to ensure open liquidity and resource discovery. Coordination is managed through smart contracts that automate resource allocation via market-based mechanisms.

\item Based on the constructed LACNets collaborative framework, our scheme can improve task processing latency, resilience, and resource utilization. This framework achieves higher overall network efficiency by aggregating computility and distributing it where it is needed. We will illustrate this point through a case study.

\end{itemize}

The rest of this article is organized as follows. Section \ref{sec-II} first clarifies the fundamental technologies, namely, blockchain, NFTs, and RWAs. We then describe the architecture and model design of RWA-powered LACNets in Sections \ref{sec-III} and \ref{sec-IV}, respectively. Next, a case study set in a smart city is presented in Section \ref{sec-V} to demonstrate how a LACNet might operate for urban drone logistics, with discussion of performance simulation results, such as latency, trust, and efficiency. Moreover, we discuss future research directions in Section \ref{sec-VI}. Finally, Section \ref{sec-VII} concludes our study.

\section{Blockchain, NFT, and RWA Technologies: Distinctions and Synergies} \label{sec-II}

This section summarizes the similarities and differences among blockchain, NFT, and RWA in Table \ref{tab1}. We will then elaborate on each of them separately.

\begin{table*}[!t]
\centering %
    \centering
    \caption{Comparison among Blockchain, NFT, and RWA}
    
    \renewcommand{\arrayrulewidth}{0.8pt} 
    \renewcommand{\tabcolsep}{10pt} 
    

    \definecolor{lightgray}{gray}{0.9} 
    \definecolor{softblue}{RGB}{220,226,235} 
    \definecolor{softgreen}{RGB}{220,230,220} 
    \definecolor{softbeige}{RGB}{245,240,225} 
    
    {\fontsize{8}{10}\selectfont 
     
    \begin{tabular}{m{1.5cm}||m{3.2cm}|m{3.2cm}|m{3cm}|m{3cm}} 
        \hline
        \rowcolor{lightgray} 
         \textbf{Technology} & \textbf{Definition}  & \textbf{Core Function} & \textbf{Applications} & \textbf{Role in LACNets}\\ 
        \hline

        \rowcolor{softblue}
         Blockchain & A Shared, immutable digital ledger technology distributed across a network. & Providing a tamper-proof, decentralized record of transactions and data trust without intermediaries. & Cryptocurrencies, healthcare records, finance, voting systems, etc.&Serving as the trust backbone for LAENets by securing flight data and transactions.\\    
         \hline

        \rowcolor{softgreen}
          NFT & 
Unique digital token on a blockchain that certifies ownership and authenticity of a specific asset. &  Representing and tracking one-of-a-kind assets by providing on-chain proof of ownership and provenance. &Digital art collections, collectibles, and in-game virtual goods. & Providing unique digital identifiers for aircraft, enabling tamper-proof certificates of ownership and usage history on-chain.\\
          
            \hline

        \rowcolor{softbeige} 
         RWA & Tokenization of physical or traditional assets into digital tokens on a blockchain. &  Bridging physical assets to blockchain by converting ownership rights into on-chain tokens. &Tokenized real estate properties, commodity-backed stablecoins, on-chain stocks/bonds.&Integrating aircraft into the broader economy by tokenizing real-world elements of the network.\\

        \hline   
     
    \end{tabular}}
    
    \label{tab1}
\end{table*}

 \subsection{Blockchain}
 
\begin{figure}[!t]
\centering
 \includegraphics[width=3.4 in]{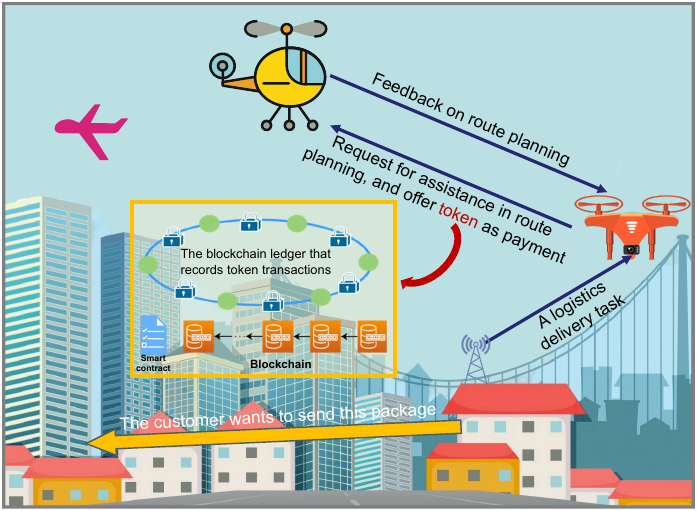}
   \caption{An example of aircraft collaboration in LACNets. When consumers want to send a package to a specific location, they will submit a request to Aircraft $A$. However, this aircraft is lacking the necessary computing resources and thus needs another Aircraft $B$ to collaborate in planning the flight path. In return, Aircraft $A$ will give $B$ appropriate tokens. This transaction will be recorded in the blockchain ledger.}
\label{fig1}
\end{figure}

Blockchain is the foundational technology enabling our approach. It is a distributed ledger maintained by a network of nodes, providing a tamper-proof record of transactions and states. Its key features, decentralization, immutability, transparency, and automation via smart contracts, address the trust and coordination needs of LAENets \cite{luo2024symbiotic}. By using a blockchain, we ensure that each aircraft’s contributions or resource usage are logged on an immutable ledger and that resource-sharing agreements are enforced by smart contracts rather than by trusting a single server. For instance, if Aircraft $A$ computes a task for Aircraft $B$, a smart contract can automatically transfer a token from $B$ to $A$ upon verified completion, without requiring a broker, as shown in Fig. \ref{fig1}. Blockchain therefore acts as the “glue” for multi-stakeholder drone networks, handling identity, access control, and transaction settlement in a verifiable way. It also improves security.

Moreover, not all blockchains are the same. We distinguish permissionless vs. permissioned blockchains for this application \cite{liu2024blockchain}. Permissionless blockchains like Ethereum offer a globally open platform with well-established token standards. They enable permissionless access, so anyone can mint or trade tokens representing drone assets. This could be useful if we envision a marketplace for aircraft computing services. But permissionless chains face scalability and privacy issues. Ethereum’s traditional Proof-of-Stake consensus is slow and has low throughput, and all transaction data is publicly visible\footnote{https://ethereum.org/en/}. In industrial or government-run UAV networks, unrestricted openness might not be ideal for security and regulatory compliance. While permissioned blockchains like Hyperledger Fabric offer an alternative: a private or consortium ledger where only authorized parties can participate\footnote{https://github.com/hyperledger/fabric}. It allows creating private channels so that, for example, drones in one city’s network only share data among themselves and a regulator, ensuring confidentiality. Hyperledger’s consensus is more scalable for a known set of nodes, which can be advantageous for the real-time constraints of aircraft coordination \cite{luo2025wireless}. In summary, public chains excel at open interoperability and liquidity of tokens, whereas private chains excel at controlled collaboration and performance.

A LACNet might even use a hybrid approach: a permissioned blockchain for real-time operational coordination and a permissionless blockchain for higher-level asset trading or settlement.

\subsection{Non-Fungible Token (NFT)}

NFTs are a type of digital token that represents unique assets\footnote{https://scand.com/company/blog/nft-standards-erc-721-vs-erc-1155-and-beyond/}. On Ethereum, the ERC-721 standard defines NFTs as one-of-a-kind tokens that cannot be exchanged one-for-one with another, unlike identical cryptocurrencies. Each NFT has a unique identifier and often metadata describing the asset. NFTs rose to fame through digital art and collectibles, but their utility goes far beyond that. They are an ideal tool for representing physical assets or rights on a blockchain because they can encode ownership and uniqueness. In our context, an NFT could represent a specific aircraft or a specific resource unit provided by an aircraft. For example, we can mint an NFT for each aircraft’s onboard compute module, with metadata describing its processor type, location, owner, etc. Owning that NFT, or a fraction of it, could entitle one to utilize that aircraft’s computing power under certain conditions. Alternatively, NFTs could represent service contracts or time slices. For example, an NFT that signifies “10 minutes of computing on Aircraft $A$ at low altitude in Shanghai”. Unlike fungible tokens, NFTs ensure that each unit is distinct and traceable, which fits the reality that each drone’s capabilities and availability are distinct.

Furthermore, Ethereum’s ERC-1155 multi-token standard adds flexibility by allowing one contract to manage multiple token types, including both fungible and non-fungible units. An ERC-1155 token can be semi-fungible. For instance, a batch of identical usage coupons, each worth 1 compute task on any drone of a certain class, could be issued. Once redeemed, a coupon could convert into a unique record of the completed service. This standard is gas-efficient for batch operations, important for IoT scale. In LACNets, we might leverage ERC-1155 to handle large numbers of short-term resource tokens.

\subsection{Real-World Asset (RWA)}

It is important to note that NFTs are a technical mechanism, whereas RWAs are a broader concept of linking digital tokens to real, off-chain assets or income streams. An NFT can be used to implement an RWA by encoding claims on a physical asset. But not all NFTs are RWAs, because some represent purely digital items with no physical counterpart. Also, not all RWAs must be NFTs, since some could be fungible, like tokenized gold or real estate shares. What RWAs add is the requirement of attestation and trust bridges between the digital token and the physical world \cite{xia2025exploration}. For tokenized aircraft computility, an oracle or attestation system is needed to ensure that when a token is issued or a smart contract triggers an aircraft task, the physical aircraft indeed provides the service. For aircraft, modules like secure elements or remote IDs might serve to cryptographically sign data about their performance, which the blockchain can verify.

The synergy between blockchain, NFTs, and RWA tokenization in our approach can be summarized as follows:

\begin{itemize} 

\item Blockchain provides a decentralized ledger and execution platform to automate and secure resource-sharing transactions in the drone network.

\item NFT standards provide the language/protocol for representing unique aircraft resources and usage rights on that blockchain, enabling interoperability and standard interfaces.

\item RWA frameworks ensure that these tokens are grounded in reality: Each token is tied to a verifiable physical asset or service. In our case, the RWA is the computing power of an aircraft, which is a somewhat abstract asset but nonetheless tangible. It is backed by physical hardware and energy. By tokenizing compute, we effectively create a new asset class: computing as an investable and tradeable commodity. Aircraft owners can monetize idle compute cycles securely.

\end{itemize}

\subsection{Lessons Learned}

NFTs provide the token container, blockchain provides the rails, and RWA tokenization provides the real-world value linkage. Together, they allow LAENet participants to convert what was previously a closed, local resource, such as onboard computing, into a part of a broader digital economy. In the next section, we explain how this is implemented in a network architecture and how standards like ERC-721/ERC-1155 come into play in building LACNets.

\section{Architecture of RWA-Powered LACNets} \label{sec-III}

\begin{figure*}[!t]
   \centering
   \includegraphics[width=5.5 in]{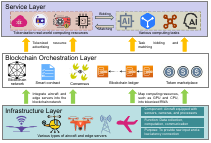}
   \caption{Three-layer architecture of RWA-Powered LACNets. The infrastructure layer consists of hardware devices that provide computing resources for LACNets or request computing services, including aircraft and edge servers. The blockchain layer builds a trusted and secure environment for LACNets and maps computing resources to RWAs; The service layer will make reasonable arrangements based on the idle RWAs and computing service requests.}
   \label{fig2}
\end{figure*}

To realize LACNets, we propose an architecture that layers a blockchain-based resource management system on top of the physical LAENet. The architecture comprises several key components, illustrated conceptually in Fig. \ref{fig2}.

\subsection{Infrastructure Layer}

The lowest infrastructure layer comprises individual UAVs or eVTOLs equipped with sensors, cameras, and onboard processors. Each aerial node collects real-world data and executes low-level tasks, e.g., sensing, communication, and basic computing. Wireless radios, such as Wi-Fi, LTE, and 5G, enable data exchange and coordination among UAVs without relying on a central controller. In this layer, each drone also performs blockchain-related primitives, that is, signing transactions, using its onboard compute. Collectively, this layer provides the raw inputs and implements low-latency links needed for higher-level coordination.

\subsection{Blockchain Orchestration Layer}

Above the physical devices sits the Blockchain Orchestration Layer, which mediates trust and coordination via a distributed ledger and smart contracts. Every aircraft, edge server, or operator agent in this layer acts as a blockchain node that records validated transactions, e.g., resource requests and token transfers, in an immutable ledger. Consensus protocols ensure that all honest nodes agree on the system state. This layer provides the tokenization fabric. The eVTOLs, sensors, compute units, or other equipment are treated as RWAs, represented as on-chain tokens, making them liquid and tradable. Smart contracts encode the rules of task allocation and payment. For example, a contract may automatically execute resource auctions or verify service delivery in exchange for tokens. By design, the blockchain layer guarantees tamper-resistant record-keeping and transparent exchanges. Cryptographic hashing and multi-party consensus prevent data tampering, while smart contracts automate task dispatch and settlement, reducing human intervention and improving fairness. The result is a decentralized marketplace in which aircraft and edge nodes interact according to provable protocols.

\subsection{Service Layer}

At the top sits the Service Layer, consisting of high-level applications, task orchestrators, and edge computing services. Here, user requests and computational tasks are generated and scheduled. Edge servers advertise available computing power, storage, or sensing resources as tokenized assets. Similarly, aircraft as service providers announce capabilities or bids for tasks on the blockchain. The Service Layer implements the economic and scheduling logic. It matches resource supply and demand via formal market mechanisms, such as game theory or auctions, governs how tokens are issued or burned, and interfaces with external networks. In practice, a smart-contract-managed marketplace at this layer collects bids and asks from aircraft and tasks, and computes allocations according to a defined protocol \cite{wu2024dynamic}.

\section{Model Design of RWA-Powered LACNets} \label{sec-IV}

This section will introduce the collaborative mechanism and market mechanism in sequence. They jointly support the implementation of the architecture described in the previous section.


\subsection{Coordination Mechanism}
Coordination among aircraft, blockchain agents, and edge systems is realized through smart contracts and consensus-driven protocols. Conceptually, each Aircraft $A_i$ is both a mobility agent and a blockchain agent. It periodically broadcasts its status, e.g., location, available compute, on the communication channel, and can submit requests: “$A_i$ needs 100 CPU cycles”, to the blockchain ledger. The blockchain layer interposes, and a smart contract will automatically trigger when a request appears. For example, consider a drone $A_i$ needing to compute for a task. $A_i$’s request is posted as a transaction. A contract then initiates a bidding phase among eligible computing nodes $C_j$, collecting their bids in tokens and resource quotes. After the auction, the contract selects the winning $C_j$, finalizes the resource trade, and issues tokens to $C_j$ while debiting $A_i$. The transaction is logged on-chain, and $C_j$ begins execution. In this way, each aircraft and edge system coordinates via an explicit contract-driven workflow. More formally, we view this as a distributed multi-agent control problem. Aircraft $A_i$ has utility $U_i(\cdot)$ based on task completion and token expenditure. Computing server $C_j$ has utility $U_j(\cdot)$ based on token revenue minus energy cost. The smart contract enforces protocol rules, e.g., payment upon completion. Consensus ensures that all participants observe the same sequence of trades. Because the blockchain is open to the participating set of nodes, malicious actors are limited by encryption and majority consensus: only requests and bids from authenticated keys $pk_{A_i},pk_{C_j}$ are accepted. 

\subsection{Market Model for Resource Tokenization}
We model resource exchange and task allocation as a formal market using game-theoretic and auction-theoretic constructs. In one view, consider a hierarchical Stackelberg game. At stage 1, computing providers $C_j$ act as leaders setting unit prices $p_j$ for computing resource tokens, and Aircraft $A_i$ act as followers choosing quantities $x_{ij}$ of resources to buy. Each leader $C_j$ aims to maximize profit $U_j = p_j\sum_i x_{ij} - O_j(\sum_i x_{ij})$, where $O_j(\cdot)$ is the overhead function. Each Aircraft $A_i$ maximizes $U_i = V_i({x_{ij}}) - \sum_j p_j x_{ij}$, where $V_i$ is the value of completed tasks minus energy cost. Under standard assumptions (convex costs, concave utilities), a unique Stackelberg equilibrium exists. In practice, $C_j$ first sets $p_j$, then each aircraft selects $x_{ij}$ accordingly.

 At stage 2, after resources are provisioned, aircraft may further allocate tasks to end users. This can be formulated as an auction. Aircraft post packages of computing service (token bundles), and user agents bid for these via on-chain auctions. For instance, a sealed-bid or Vickrey auction can be implemented by a smart contract \cite{bag2019seal}, so that each user’s bid reveals their willingness to pay for a token bundle. The contract then allocates service to the highest bidders and settles payments in RWA tokens. This two-tier (leaders and followers) game, combined with auctions, ensures that resources are priced according to supply-demand balance, and tasks are assigned to drones that value them most.
 
 Alternatively, one can model the entire market as a double auction: drones submit asks (for tokens) and users submit bids (for tasks) simultaneously. The smart contract then clears the market by finding a set of matches that maximizes social welfare or budget balance. This decentralized auction can be proven to be incentive-compatible under certain mechanism design constraints. In all cases, the formal model provides equations for agents’ payoffs and market equilibrium conditions, which can be solved analytically or algorithmically to set optimal prices and allocations \cite{chai2020hierarchical}.

\begin{figure}[!t]
\centering
 \includegraphics[width=3.2 in]{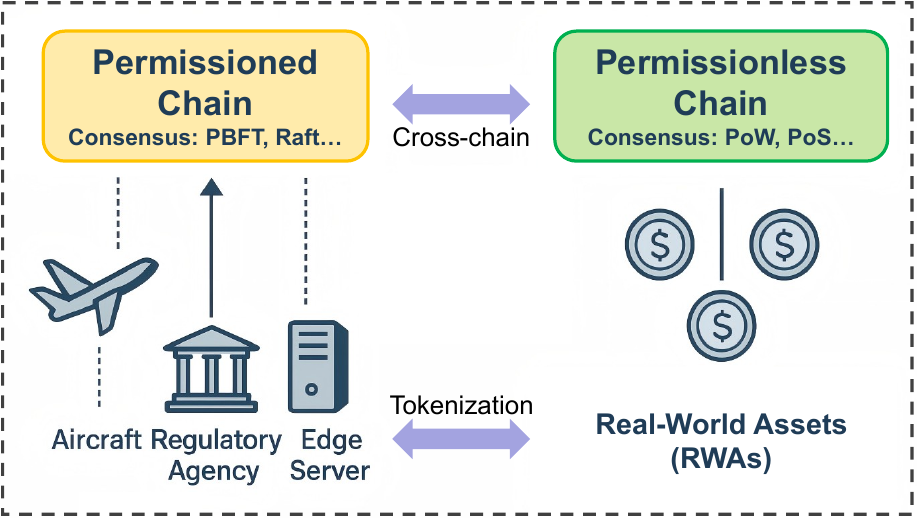}
   \caption{Collaboration between the permissioned chain and the permissionless chain. Through tokenization, the aircraft computing resources are mapped to RWAs and managed by a permissionless blockchain; Aircraft, edge servers, and city administrators are registered through the permission chain. The two chains are synchronized and anchored through the cross-chain.}
\label{fig3}
\end{figure}

\subsection{Blockchain Design}

In LACNets, two blockchain models can be adopted, as shown in Fig. \ref{fig3}. The permissioned chain limits participation rights to authorized entities only. For example, all aircraft, regulatory agencies, and edge servers can form an alliance. This controlled environment is usually more suitable for aviation systems where security is of paramount importance. The other is the permissionless blockchain, where any node can join and participate in the consensus process. It is targeted at tokenized RWAs. This approach maximizes openness and decentralization, significantly promoting the open liquidity of RWAs.

The overall approach is to incorporate identity and permission control within the permissioned chain, implementing device whitelisting and encrypted data storage, while placing globally verifiable RWAs on the permissionless chain for resource discovery and settlement. Finally, the two are securely bonded via a cross-chain mechanism. If a device’s private data does not match the public chain data, anchoring records can be used for reconciliation \cite{jiang2025towards}. In case of disputes, the private chain history can be replayed to obtain evidence.

\section{Case Study: A LACNet for Urban Logistics} \label{sec-V}

\subsection{Case Scenario}

To illustrate the LACNet, we present a scenario involving a consortium of stakeholders: a logistics company operating delivery drones, a municipal government overseeing air traffic and data, a tech startup providing an AI analytics service that needs edge computing, and an eVTOL operator providing aerial taxi services. All these parties are part of the LAENets, focusing on urban logistics and transport.

\begin{figure}[!t]
\centering
 \includegraphics[width=3.4 in]{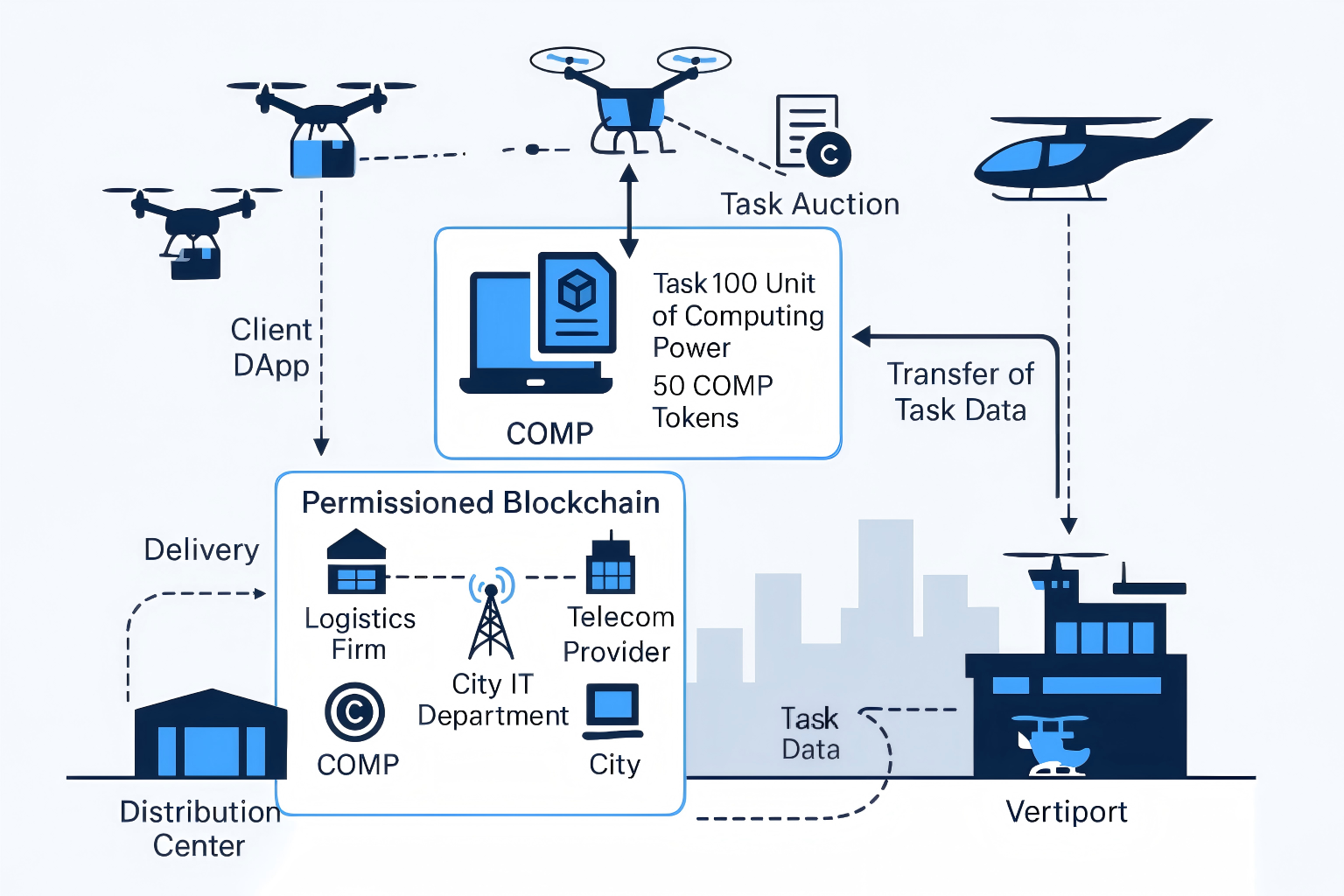}
   \caption{A scenario case of LACNets. A large number of delivery drones and eVTOLs operate in the city and generate idle computing power. The permissioned chain registers their identities and resources. The computing demanders outsource computing tasks to idle aviation nodes through COMP tokens and auction contracts on the permissionless chain. }
\label{fig4}
\end{figure}

Every day, hundreds of delivery drones fly across the city, transporting packages between distribution hubs and neighborhoods, as shown in Fig. \ref{fig4}. They are equipped with cameras for navigation and obstacle avoidance, and onboard computers that can run basic autonomy algorithms. Separately, the city’s eVTOL air taxis ferry passengers along designated routes, also with powerful computing for navigation and safety systems. During operations, there is spare computing capacity, for example, when cruising or hovering, a drone’s CPU/GPU is not fully utilized. Meanwhile, the AI startup needs to process large amounts of video data at the edge (e.g., analyzing foot traffic or vehicle counts from street cameras) to provide real-time insights to businesses and city planners. Instead of building its own fixed edge server network, the startup opts to offload tasks to any available aerial node in the LACNet, essentially using idle aircraft as airborne computers.

The consortium sets up a permissioned blockchain network that includes nodes run by the logistics firm, the eVTOL company, the city IT department, and a neutral telecom provider. Each drone and eVTOL is registered on this blockchain via its ground station when it leaves the hangar. For simplicity, the consortium defines a fungible token called COMP that represents a unit of computation (1 GFLOP of computing). Drones and eVTOLs earn COMP tokens by performing approved computations for the network. Conversely, the AI company and others who need computing power buy COMP tokens to pay for these services. The COMP token is implemented as an ERC-721/1155-like asset on the permissionless chain. Each token’s issuance is tied to actual compute performed. One could think of COMP tokens as analogous to how electricity is traded in energy grids, but here it’s computing power being metered and traded.

As drones launch to deliver packages, they check in to the blockchain. A smart contract records their current load and available compute capacity. Suppose the AI startup needs to run a computer vision algorithm on 100 video frames near real-time in a certain district. Through a client DApp (decentralized app), it creates a task request on the blockchain: 100 units of computing power are required within 2 minutes, with a proposed payment of 50 COMP tokens, that is, a willingness-to-pay price of 0.5 tokens per unit. This task is posted to a task auction contract visible to all aircraft in that district. Two delivery drones and one idle eVTOL see the task. Their onboard agents evaluate if they can fulfill it. One drone is too busy with a delivery and ignores it. Another drone has some slack and bids to process 30 of the frames for 0.4 tokens each. The eVTOL, which is currently charging on a rooftop vertiport and has a very powerful computer, bids to process all 100 frames for 0.3 tokens each. It is mostly idle and willing to earn anything. The smart contract collects bids for a few seconds, then assigns the job to the eVTOL as it offered the lowest cost and can do it fully. The task data, namely video frames, is delivered to the eVTOL either directly via a device-to-device link or through an edge server. The contract might provide a secure hash for the data that only the eVTOL can access. The hash is an output of an oracle that bridges to an off-chain data exchange service.

\subsection{Simulation Setup}

The simulation is configured within a $2×2$ km urban environment, populated by a variable number of aerial nodes ranging from $20$ to $100$. This fleet consists of $80\%$ UAVs, modeled after the DJI Matrice M300 RTK with a maximum speed of $23$ m/s  and a flight altitude of $100-150$ m\footnote{https://www.dji.com/sg/support/product/matrice-300}, and $20\%$ more powerful eVTOLs with an assumed maximum speed of $45$ m/s. All nodes communicate over a simulated 5G network, featuring $5$ Gbps bandwidth and an average latency of $10$ ms\footnote{https://advexure.com/blogs/news/the-role-of-5g-in-expanding-drone-capabilities}. For the RWA-LACNet, we simulate the blockchain with a maximum throughput of $1,000$ TPS, a $2$-second block timeout, a block size of $10$ transactions, and a baseline transaction latency of approximately $1.1$ seconds. Onboard computational power is set to $1$ TFLOPS for UAVs, akin to the NVIDIA Jetson TX1\footnote{https://www.rs-online.com/designspark/a-first-look-at-the-nvidia-jetson-agx-orin}, and $5.3$ TFLOPS for eVTOLs, based on the NVIDIA Jetson AGX Orin\footnote{https://www.techpowerup.com/gpu-specs/jetson-agx-orin-64-gb.c4085}. The simulation involves video analysis tasks arriving at a variable Poisson rate of $10$ to $200$ tasks per minute. Each task requires the transfer of $25$ MB of data and a computational load of $2$ TFLOPs\footnote{https://mlsys.org/Conferences/2019/doc/2019/167.pdf}. This simulation runs on a server equipped with a 96-core Intel(R) Xeon(R) Gold 5220R CPU @ 2.20 GHz with 1 TB of memory.

Our comparison schemes with RWA-LACNet are the Centralized Task Allocation (CTA) scheme \cite{han2024collaborative} and the decentralized task allocation scheme \cite{wu2024dynamic}. The former represents the traditional command and control approach, where a single, powerful Ground Control Station (GCS) is responsible for all decisions. All aircraft report their status to the GCS. When a new task arrives, the GCS runs a centralized optimization algorithm to allocate the task to the most suitable airborne node. The latter eliminates the central controller and implements a decentralized task allocation method based on auctions: the Consensus-Based Bundle Algorithm (CBBA). In this model, each aircraft is an autonomous agent that can build a bundle of the tasks it can perform and provide bids for each task. Through consensus negotiation, aircraft update each other on winning bid information for each task, resolve conflicts, and continue until the allocation scheme converges.

\subsection{Simulation Results}

This part presents the results of the simulation, organized around three key performance indicators: task processing latency, resilience to attacks, and computility utilization.

\begin{figure*}[!t]
   \centering
   \includegraphics[width=6.5 in]{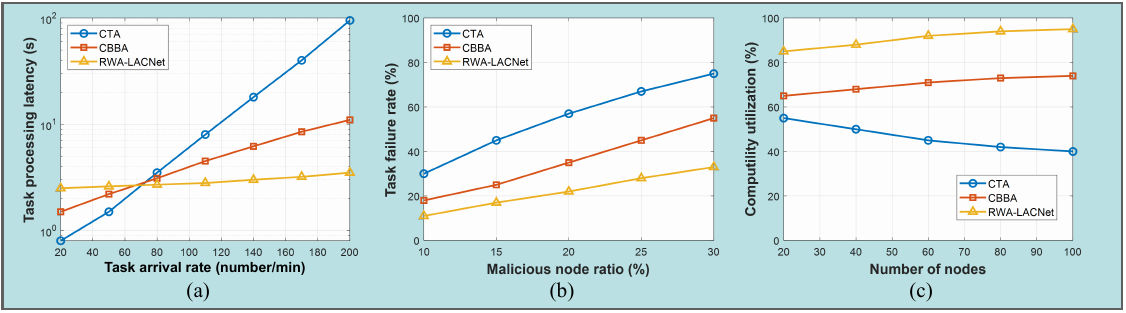}
   \caption{Simulation results. (a) Task processing latency; (b) Task failure rate; (c) Computility utilization.}
   \label{fig5}
\end{figure*}

Fig. \ref{fig5} (a) compares task processing latency across the three schemes under different task arrival rates. Among them, CTA shows the lowest latency at extremely low loads. However, as the task arrival rate increases, the GCS becomes a performance bottleneck, causing queuing delays to rise sharply and overall latency to grow exponentially. Additionally, the latency performance of the CBBA model is intermediate. But as the network becomes busy, the communication overhead required to reach consensus increases, leading to higher latency. While our RWA-LACNet model has the highest latency at low task arrival rates, due to block creation time, its latency curve remains flat, scaling nearly linearly and demonstrating excellent scalability. This is due to the distributed auction mechanism and the dual design using both permissioned and permissionless chains. Once an aircraft joins the LACNet, it can quickly trade tokenized RWAs.

Fig. \ref{fig5} (b) shows the trend of task failure rates for the three schemes as the proportion of malicious nodes increases. Among them, the CTA’s task failure rate is the most vulnerable, rising significantly as the proportion of malicious nodes increases. As the central decision-maker, the GCS’s allocation algorithm relies entirely on the accuracy of data reported by each node. Additionally, CBBA shows an intermediate level. With its consensus mechanism, it can cope with interference from a small number of malicious nodes. Our RWA-LACNet model demonstrates the strongest resilience, with the most gradual increase in task failure rate. Its failure rate is approximately equal to the proportion of tasks won by malicious nodes that are abandoned. Its resilience stems from algorithmic trust and economic incentive mechanisms: it does not trust participants but relies on cryptographically verified, immutable smart contract code. Malicious behavior incurs direct economic penalties, significantly reducing the willingness to attacks and their impact.

Fig. \ref{fig5} (c) compares the computility utilization rates of the three schemes. Among them, CTA has the lowest utilization rate. The GCS’s bottleneck effect prevents computational tasks from being dispatched quickly enough to keep all nodes busy, leaving high-performance eVTOLs potentially idle. Additionally, CBBA has a moderate utilization rate, better than CTA. However, due to weak incentives, nodes may not voluntarily bid for tasks unless programmed to do so. Our RWA-LACNet has the highest utilization rate: economic incentives drive idle nodes to participate in auctions, forming a more efficient market that matches computing supply and demand. Powerful but idle eVTOLs become highly sought-after resources.

Overall, RWA-LACNet represents a paradigm shift in distributed resource orchestration by integrating a blockchain-based economic layer. It is not merely a technical solution but a self-regulating micro-economy. By tokenizing computing power into tradable RWAs, it successfully addresses critical issues of trust and efficiency in the multi-stakeholder low-altitude economy.

\section{Future Research Directions} \label{sec-VI}

\subsection{AI-Orchestrated Resource Management}

While our case used relatively simple auction contracts, future systems could leverage AI for intelligent orchestration. For example, machine learning agents could predict demand and preemptively adjust token prices or allocate resources. A multi-agent AI system might manage different aspects: one agent optimizes the route and location of drones, another balances their energy use between computing tasks, and another negotiates in the token marketplace. These agents could even be implemented as on-chain AI Decentralized Autonomous Organizations (DAOs) that adjust rules on the fly. Integrating such AI decision-making with the blockchain layer can greatly improve efficiency. However, challenges include the complexity of real-time learning and ensuring the AI’s decisions are transparent and fair in an autonomous market.

\subsection{Edge AI Offloading and Collaborative Computing}

LACNets effectively create a fog computing layer, which raises research questions about how to optimally partition tasks among aircraft, edge servers, and the cloud. In our architecture, some tasks were fully offloaded to the aircraft. But we could consider split computing, where a heavy AI model is partitioned, with some layers running on the drone and the rest on an edge server or the cloud. Such co-inference can reduce latency and energy consumption. How to tokenize and reward partial contributions in such cases is an interesting problem. Relatedly, aircraft are energy-constrained, and heavy computing drains batteries. Future LACNets might integrate energy-aware scheduling. For example, an aircraft nearing low battery should trade away tasks to others or to ground edge nodes, even if it’s capable. Tokens might even incorporate energy costs. An exciting direction is coupling renewable energy grids with these computility networks. Aircraft charging at solar-powered stations could perform computational tasks when solar energy is abundant, essentially converting excess electricity into useful computation.

\subsection{Cross-Jurisdictional and Policy for RWAs}

A major non-technical hurdle for tokenizing real aircraft resources is regulatory acceptance. RWAs blur the lines between financial instruments and physical asset leasing. Each jurisdiction has different laws on tokenized assets, securities, data handling, and aviation. Future work must engage with policymakers to develop frameworks that recognize RWA tokens as legal contracts for service. This might involve standardizing what an RWA token entails. This domain will need to craft rules that both foster innovation and protect public safety and privacy. Fortunately, the data immutability of blockchains can aid regulation. For example, regulators could be given a node that monitors all RWA transactions to flag unauthorized usage or enforce taxes on drone services automatically through smart contracts.

\section{Conclusion} \label{sec-VII}

This article introduced RWA-LACNet, a novel framework designed to orchestrate the underutilized computing resources of aerial vehicles in LAENets. By leveraging RWA tokenization, a hybrid blockchain architecture, and smart contract-driven auction mechanisms, RWA-LACNet transforms siloed, onboard computing power into a liquid, tradable utility. This approach establishes a decentralized marketplace that significantly enhances resource sharing and economic efficiency in multi-stakeholder aerial environments. Our simulation-based case study in an urban logistics scenario validated the superiority of RWA-LACNet over traditional centralized and decentralized consensus models, demonstrating superior scalability in task processing, stronger resilience against malicious attacks, and higher overall resource utilization. These advancements will be critical in paving the way for the broader real-world deployment of a secure, efficient, and economically vibrant low-altitude economy.

\bibliographystyle{IEEEtran}
\bibliography{IEEEabrv,mylib}



\vspace{3em}
\vspace{3em}

{\footnotesize

}

\vfill

\end{document}